\documentclass[a4paper, 10pt, conference]{ieeeconf}      

\IEEEoverridecommandlockouts                              
\usepackage{graphics} 
\usepackage{epsfig} 
\usepackage{rotating}

\title{\LARGE \bf
Overlapping Probabilities of Top Ranking Gene Lists,
Hypergeometric Distribution, and Stringency of Gene Selection Criterion
}

\author{Wen Fury, Franak Batliwalla, Peter K. Gregersen, and Wentian Li
\thanks{W. Fury is a Senior Bioinformatics Scientist at Regeneron Pharmaceutical, Inc.
	Tarrytown, NY 10591, USA.
        {\tt\small wen.fury@regeneron.com}}%
\thanks{F. Batliwalla, P.K. Gregersen, and W. Li are Research Scientists
with the Robert S Boas Center for Genomics and Human Genetics, 
Feinstein Institute for Medical Research, North Shore LIJ Health System,
	Manhasset, NY 11030, USA
        {\tt\small fb@nshs.edu},
        {\tt\small peterg@nshs.edu},
        {\tt\small wli@nslij-genetics.org}}%
}

\begin{document}

\maketitle
\thispagestyle{empty}
\pagestyle{empty}

\begin{abstract}

When the same set of genes appear in two top ranking gene lists in
two different studies, it is often of interest to estimate
the probability for this being a chance event. This overlapping
probability is well known to follow the hypergeometric
distribution.  Usually, the lengths of top-ranking gene lists 
are assumed to be fixed, by using a pre-set criterion on, e.g.,
$p$-value for the $t$-test. We investigate how overlapping probability
changes with the gene selection criterion, or simply, with the
length of the top-ranking gene lists. It is concluded that 
overlapping probability is indeed a function of the gene list 
length, and its statistical significance should be quoted in 
the context of gene selection criterion.

\end{abstract}

\section{INTRODUCTION}

One of the most common tasks in microarray analysis 
is to identify a list of genes that are differentially
expressed under two conditions, such as being affected by
a disease vs. normal, before vs. after a medical
treatment, and one vs. another disease subtype. The
number of genes on the top-ranking list is
usually much smaller than the total number of genes
on the chip, $n$. If the same type of microarray chip is used for 
two different studies (e.g. disease-A vs. control, 
and disease-B vs. control), two differentially
expressed gene lists can be obtained, with $n_1$ and
$n_2$ genes. Researchers often find the same genes
appear in both lists and hypothesize that these common
genes are involved the etiology of both diseases.

However, for such a hypothesis to be convincing,
one has to first estimate the probability for 
overlapping genes by chance alone. In other words,
if two lists of genes are selected out of $n$ genes 
randomly, we would like to calculate the probability
for $m$ genes in common in the two lists,
with the lengths of the two lists being $n_1$ and $n_2$.
This overlapping probability is known to follow the
hypergeometric distribution \footnote{Despite certain
similarity, this problem is not the birthday problem 
-- the probability for two people in a room to 
have the same birthday.}. The name hypergeometric
distribution was first used in \cite{hyper}, and
was popularized by its role in Fisher's exact
test \cite{fisher}.

In microarray analysis, overlapping probability and
hypergeometric distribution mainly appear in testing
the enrichment of genes in certain functional
category \cite{tavazoie, draghici, fino, hosack,  
boorsma, curtis, mao, tian}. In this application,
the first list is the top-ranking differentially
expressed genes, and a gene selection process is
involved. The second list is nevertheless given: 
$n_2$ genes are known to be in a pathway, a 
member of a protein family, described by a gene ontology term,
etc. One asks the question on chance probability
for $m$ out of $n_1$ selected genes to be in 
a given pathway, a protein family, and describable 
by a gene ontology term.  Fixing $n_2$ or not is the 
main difference between their application and ours.

When a different gene selection criterion is used,
the number of genes in the two top-ranking lists
of two studies ($n_1$ and $n_2$) will also change.
Because the stringency of a gene selection criterion
is always adjustable and to some extent arbitrary,
we would like to examine whether these changes will
affect the overlapping probability. At two
extreme situations, very small $n_1 = n_2 \approx 1 $
and very large $n_1=n_2 =n$, it is clear that
the number of overlapping genes is $m=0$ and $m=n$.
These $m$ values appear 100\% of the times, so
the corresponding $p$-value is equal to 1, i.e.,
not significant. For intermediate $n_1 \approx n_2$
values, it is not clear what the overlapping
probability and significance will be, and it is
the topic of this abstract.

\section{HYPERGEOMETRIC DISTRIBUTION AND OVERLAPPING P-VALUES}

Given integers $n$, $n_1$, $n_2$, $m$ 
($ \max(n_1, n_2) \le n$ and $m \le \min(n_1, n_2$) ), the hypergeometric
distribution is defined as
$$
P(m) =\frac{ C(n_1, m ) C(n-n_1, n_2-m )}{ C(n, n_2) }
= \frac{ \left( \begin{array}{c} n_1 \\ m \end{array} \right)
\left(  \begin{array}{c} n-n_1 \\ n_2-m  \end{array} \right)}
{ \left( \begin{array}{c} n \\ n_2 \end{array}  \right) }
$$
where $C(n, m)$ is the number of possibilities of choosing
$m$ objects out of $n$ objects: $C(n, m)= n!/[m! (n-m) !] $.

When $n_1$ genes are randomly chosen from the total of
$n$ genes, and another random sampling leads to $n_2$
genes, the probability that the two lists of genes have
$m$ in common is exactly the hypergeometric probability
$P(m)$. This can be proven by the following steps:
1) The total number of possible choices for the two
lists of genes is $C(n, n_1) \cdot C(n, n_2)$.
2) There are $C(n, n_1)$ possibilities for choosing the first
list.
3) Among the $n_1$ genes in the first list, there are
$C(n_1, m)$ possibilities  for choosing $m$ genes to
be in common with the second list.
4) In the second list, besides the $m$ genes that are in
common with the first list, the remaining $n_2-m$ genes
are chosen among the $n-n_1$ ``leftover" genes not
in the first list, thus $C(n-n_1, n_2-m)$ possibilities.
The $P(m)$ is simply (\#2 $\times$ \#3 $\times$ \#4) / \#1.
Note that $n_1$ and $n_2$ can be switched without
changing the $P(m)$ value.

It is usually more interesting to calculate the sum of
$P(m)$ for $m$'s equal or larger than the observed value
(i.e., the $p$-value):
$$
p\mbox{-value} =  \sum_{k = m}^{\min(n_1, n_2)} p(k)
= \sum_{k=0}^{\min(n_1, n_2)} p(k)
-\sum_{k=0}^{m-1} p(k)
$$
In statistical package $R$ ({\sl http://www.r-project.org/}), 
there are at least two ways to calculate the overlapping $p$-value.
The first is to use the accumulative distribution of
hypergeometric distribution, {\sl phyper(m, $n_1$, $n-n_1$, $n_2$)}:
$p$-value $= phyper(\min(n_1, n_2), n_1, n-n_1, n_2)
- phyper(m-1, n_1, n-n_1, n_2)$ if $m >0$, and
$p$-value=1 if $m=0$. The second method is to use
the  $p$-value from the Fisher's exact test on 
the following 2-by-2 table:
$$ 
\begin{array}{c|cc|c}
 & col_1 & col_2 & total \\
\hline
 row_1& m & n_1 -m & n_1 \\
row_2& n_2-m & n -n_1-n_2+m & n-n_1 \\
\hline
total & n_2 & n-n_2 & n
\end{array}
$$ 
The two approaches lead to the identical result.
 
   \begin{figure}[t]
      \centering
	\begin{turn}{-90}
	\resizebox{8.0cm}{8.0cm}{ \includegraphics{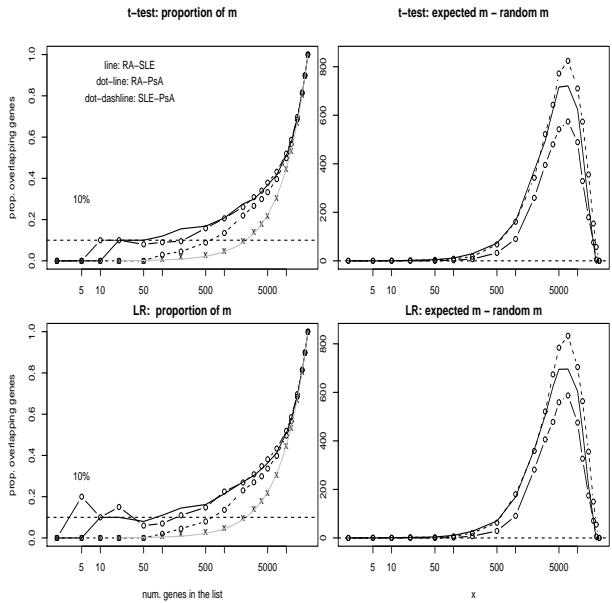} }
	\end{turn}
      \caption{First column: proportion of overlapping genes between
two top ranking gene lists for a pair of studies ($m/n_1$)
as a function of the gene list length ($n_1(=n_2)$). Top is
for gene ranking by $t$-test and bottom is for gene ranking
by logistic regression. The overlapping proportion for
two randomly shuffled lists is shown in crosses, and the line
$m/n_1 = n_1/n$ is marked. Second column: observed number
of overlapping genes ($m$) subtract the expected number
of overlapping genes ($n_1^2/n$).
	}
      \label{fig1}
   \end{figure}

\section{PROPORTION OF OVERLAPPING GENES IN A COLLECTION
OF MICROARRAY  DATASET}

In hypergeometric distribution, the number of overlapping
elements $m$ is an independent variable from the the
list lengths $n_1, n_2$. In order to get a rough idea on
how $m$ changes with the list lengths, we use three real
microarray datasets.  Theese studies concern three 
autoimmune diseases: rheumatoid 
arthritis (RA), systemic lupus erythematosus (SLE), and 
psoriatic arthritis (PsA), described in details in
\cite{ra, sle, psa}.  The number of controls (C) and patients (P)
in these three datasets are (C=39, P=46), (C=41, P=81), and 
(C=19, P=19), respectively. The total number of genes/probe-sets
is $n=$22283, and  the expression levels are log transformed.
Genes are ranked for their degree of differential expression 
which can be measured by various tests or models, such 
as $t$-test and logistic regression.

For any pair of studies, with a fixed number of top-ranking
gene lists $n_1(=n_2)$, one can count the number of overlapping genes
$m$ and the proportion $m/n_1(=m/n_2)$. Fig.\ref{fig1} (left
column) shows this proportion as a function of $n_1(=n_2)$ 
for three study-pairs (RA-SLE, SLE-PsA, RA-PsA) as well as for two ranking methods 
($t$-test and logistic regression). Similar overlapping 
proportion of two random shuffled lists is also 
indicated in Fig.\ref{fig1} as crosses.

When $n_1(=n_2)$ is small, $m$ is more likely to be zero, so
the proportion is also zero. When $n_1(=n_2)$ approaches the
total number of genes, $n$, all genes are overlapping genes,
and the proportion is 1. Fig. \ref{fig1} indeed shows these
trends at the two extreme points. In order to check
behavior in-between, we draw a reference line in Fig.\ref{fig1}
(left column) that assume a linear relationship between 
$m/n_1$ and $n_1/n$.  Most of the points on Fig.\ref{fig1} 
are above this line, and the overlapping proportion of two 
random lists is exactly on this line.

To have an idea of the absolute number of common genes
more than expected by random chance, Fig.\ref{fig1} (right
column) plots the observed $m$ subtract the expected $m_{exp}= n_1^2/n(=n_2^2/n)$
as a function of $n_1(=n_2)$. The maximum difference between
the observed and expected is reached between $n_1=5000$ and
$n_1=10000$. The difference of observed and expected $m$'s 
can be as much as 600--800.

   \begin{figure}[t]
      \centering
	\begin{turn}{-90}
	\resizebox{4.0cm}{7.50cm}{ \includegraphics{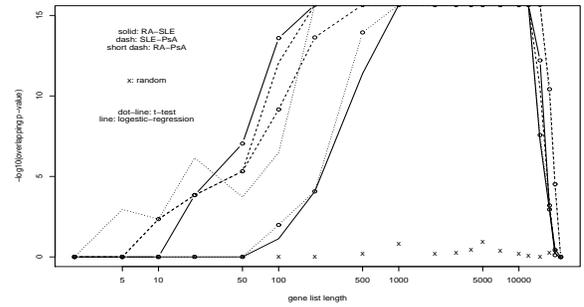} }
	\end{turn}
      \caption{
	Overlapping significance as measured by $-\log_{10}(p$-value)
where $p$-value is obtained by the hypergeometric distribution,
as a function of $n_1(=n_2)$, the number of genes in the
top-ranking gene lists. The $R$ program reports $p$-value to
be zero whenever it is lower than 2.2$\times 10^{-16}$, and
we use a ceiling of 15.65758 $=-\log_{10}(2.2 \times 10^{-16})$
in the plot.  Six lines are shown for three
study pairs (RA-SLE, SLE-PsA, RA-PsA) and two tests/models
($t$-test and logistic regression). Similar overlapping significance
for two randomly shuffled lists is also shown (indicated by crosses).
	}
      \label{fig2}
   \end{figure}

\section{OVERLAPPING SIGNIFICANCE}

The overlapping $p$-value corresponding to the $m$ counts
plotted in Fig.\ref{fig1} was calculated by the hypergeometric
distribution, and is shown in Fig.\ref{fig2}:
$y$-axis is $-\log_{10}(p$-value), and $x$-axis is
$n_1(=n_2)$. Six lines are shown for
three comparisons (RA-SLE, SLE-PsA, RA-PsA) and two
measurements of the differential expression ($t$-test and
logistic regression).  Zero $p$-values are converted to
2.2 $\times 10^{-16}$ which is the minimum value
reported by $R$ program.  Fig.\ref{fig2}  shows that 
besides the two ends ($m=n_1=n_2=0$ and $m=n_1=n_2=n$) where 
the $p$-value is 1, the overlapping significance
quickly increases with the length of top-ranking gene list
$n_1(=n_2$), and can be extremely significant when a
large number of genes are kept in the two lists
for comparison.

This result confirm our previous suspicion that overlapping
significance is a function of the gene list lengths.
If the selection of $n_1, n_2$ is arbitrary, the
overlapping significance thus calculated is also
arbitrary. It is not surprising that
overlapping significance may keep increasing
(or, $p$-value decreasing) with the increase of $n_1(=n_2)$,
because $p$-value in general depends on the sample
size. When a signal is real (true positive), $p$-value
will monotonically decrease with the sample size.
On the contrast, if a true signal is absent, the
sample size does not affect the conclusion. As
can be seen in Fig.\ref{fig2}, the overlapping significance
for two random lists does not really change with $n_1(=n_2)$.

One may argue that it is unlikely to consider
top 5000 genes as being differentially expressed,
because by a typical selection criterion (e.g. $p$-value of
$t$-test smaller than 0.01, with or without multiple
testing correction), the number of genes selected
is less than a few hundreds. However, as can be
seen in Fig.\ref{fig2},  even in the range
of 10--500, the overlapping $p$-value changes dramatically.

This pitfall of gene-list-length dependence of overlapping
$p$-values  has not been noticed before
perhaps because in other application of hypergeometric
distribution for calculating overlapping probability,
the length of the second list $n_2$ is fixed, for example,
in the study of overrepresentation of genes in
certain pathway. The number of overlapping genes $m$
is then constrained from above by $\min(n_1, n_2)$ even though
the length of the first list, $n_1$, might increase
by relaxing the gene selection criterion.

   \begin{figure}[t]
      \centering
	\begin{turn}{-90}
	\resizebox{4.0cm}{7.0cm}{ \includegraphics{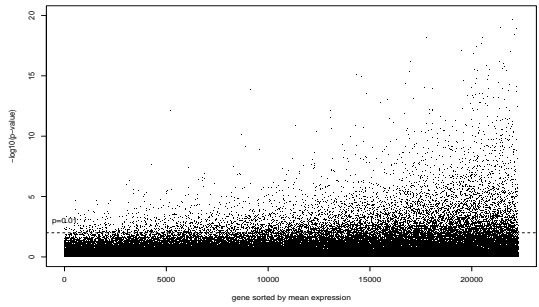} }
	\end{turn}
      \caption{The test significance ($-\log_{10}(p$-value))
from $t$-test of $n=$22283 genes sorted by the averaged expression 
level (log-transformed) across all 245 samples in 3 studies 
(RA, SLE, PsA). The three $t$-tests are for RA vs. control, SLE vs. control,
and PsA vs. control.
 	}
      \label{fig3}
   \end{figure}

   \begin{figure}[t]
      \centering
	\begin{turn}{-90}
	\resizebox{8.0cm}{8.0cm}{ \includegraphics{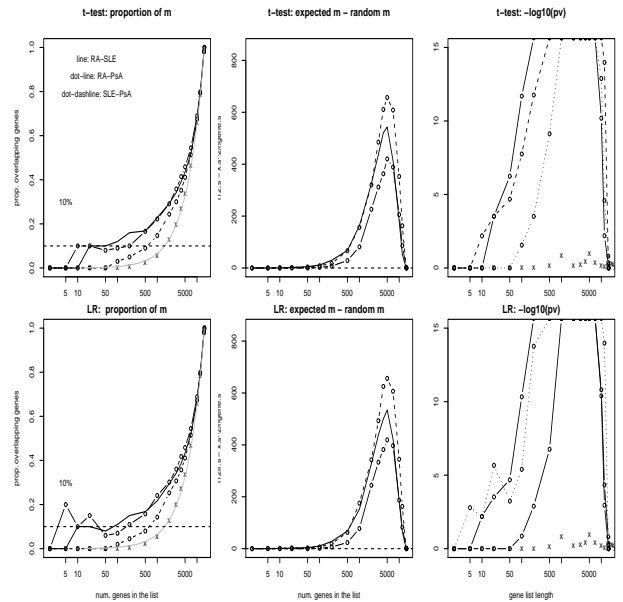} }
	\end{turn}
      \caption{Several measures of overlapping genes between
a pair of studies as a function of the number of genes included
in the top-ranking list, for the reduced dataset with 15283 genes.
First column: proportion of overlapping genes ($m/n_1$); 
second column: number of observed overlapping genes subtracting the 
number of expected ($m- n_1^2/15283$); third column: $-\log_{10}(p$-value)
by the hypergeometric distribution. First row is for lists ranked
by $t$-test result, and second row is for lists ranked by
logistic regression. 
       }
      \label{fig4}
   \end{figure}
\section{THE EFFECTS OF UNEXPRESSED GENES}

There are many genes/probe-sets on the microarray chip
that do not register much signal. Since these low-expressed
genes are lowly expressed in both control and patient
samples, they usually do not appear in the top-ranking
differentially expressed gene list.  Fig.\ref{fig3}
shows $-\log_{10}(p$-value) of each gene of 3 $t$-tests 
sorted by average expression (log-transformed)
across all 245 samples in 3 datasets (for both cases and controls). Although
we cannot use the average expression level to predict
the degree of differential expression, there is 
a general trend for low-expressed genes to rank lower in the
differentially expressed list as seen from Fig.\ref{fig3}.

We removed 7000 genes with lower overall expression across
all samples, leaving $n=15283$ genes. Figs.\ref{fig1} and \ref{fig2}
are reproduced in Fig.\ref{fig4} for the dataset with a reduced gene pool.
As in Figs.\ref{fig1} and \ref{fig2}, the observed number
of overlapping genes $m$ is much larger than the expected,
though the difference peaks at 400--600, as versus 600-800
in Fig.\ref{fig1}. The overlapping significance as measured
by $-\log(p$-value) again quickly moves up with $n_1(=n_2)$
as shown in the last column of Fig.\ref{fig4}. 

The qualitative similarity between Figs.\ref{fig1}, \ref{fig2}
and Fig.\ref{fig4} indicates that the presence of 
low-expressed genes does not affect our conclusion.

\addtolength{\textheight}{-12cm}   

\section{CONCLUSIONS AND FUTURE WORKS}

\subsection{Conclusions}

Using the hypergeometric distribution to calculate the
overlapping probability between two top-ranking differentially
expressed genes in two studies, we have shown that the
overlapping significance depends on the stringency of
gene selection criterion, or equivalently, the length
of the gene lists. This observation presents a problem
when an overlapping $p$-value is reported but the
gene selection criterion is not specified. On the other
hand, the increase of the overlapping significance
with the gene list length can be an indication that
the significant overlapping of genes is a true signal.

\subsection{Future Works}

The overlapping probability calculated here assumes the two 
top-ranking gene lists are selected from the same pool of $n$ 
genes. If the two studies are based on different chip
platforms, the two initial gene pools are not identical,
though there are perhaps certain common genes. We plan to
derive the overlapping distribution for this situation.

We also plan to study the probability for genes appearing
in three top-ranking gene lists. Although a permutation based
approach comparing multiple studies was proposed in \cite{rhode},
there is no analytic formula available.

\section{ACKNOWLEDGMENTS}

We would like to thank Prof. Richard Friedberg for suggestions.


\end{document}